\documentclass{pasj01}
\draft
\usepackage{booktabs}
\usepackage{wrapfig}
\usepackage{subfigure}

\begin{document}
\SetRunningHead{H., Arai, et al.}{Running Head}

\title{A Multi-Line Ammonia Survey of the Galactic Center Region
with the Tsukuba 32-m Telescope -- I. Observations and Data}

\author{Hitoshi \textsc{Arai},\altaffilmark{1,2}
			Makoto \textsc{Nagai},\altaffilmark{1,3}
			Shinji \textsc{Fujita},\altaffilmark{1}
			Naomasa \textsc{Nakai},\altaffilmark{1,3}
			Masumichi \textsc{Seta},\altaffilmark{1,4}
			Aya \textsc{Yamauchi},\altaffilmark{1,5}
			Hiroyuki \textsc{Kaneko},\altaffilmark{1,2}
			Kenzaburo \textsc{Hagiwara},\altaffilmark{1}
			Koh-ichi \textsc{Mamyoda},\altaffilmark{1}
			Yusuke \textsc{Miyamoto},\altaffilmark{1,2}
			Masa-aki \textsc{Horie},\altaffilmark{1}
			Shun \textsc{Ishii},\altaffilmark{1,6}
			Yusuke \textsc{Koide},\altaffilmark{1}
			Mitsutoshi \textsc{Ogino},\altaffilmark{1}
			Masaki \textsc{Maruyama},\altaffilmark{1}
			Katsuaki \textsc{Hirai},\altaffilmark{1}
			Wataru \textsc{Oshiro},\altaffilmark{1}
			Satoshi \textsc{Nagai},\altaffilmark{1}
			Daiki \textsc{Akiyama},\altaffilmark{1}
			Keita \textsc{Konakawa},\altaffilmark{1}
			Hiroaki \textsc{Nonogawa},\altaffilmark{1}
			Dragan \textsc{Salak},\altaffilmark{1,4}
			Yuki \textsc{Terabe},\altaffilmark{1}
			Yoshiki \textsc{Nihonmatsu}\altaffilmark{1}
			and
			Fumiyoshi \textsc{Funahashi}\altaffilmark{1}
}
\altaffiltext{1}{Division of Physics, Faculty of Pure and Applied Sciences, University of Tsukuba, 1-1-1 Ten-nodai, Tsukuba, Ibaraki 305-8571, Japan}
\email{hitoshi.arai@nao.ac.jp}
\altaffiltext{2}{Nobeyama Radio Observatory, 462-2 Nobeyama, Minamimaki, Minamisaku, Nagano 384-1305, Japan}
\altaffiltext{3}{Center for Integrated Research in Fundamental Science and Engineering, University of Tsukuba, 1-1-1 Ten-nodai, Tsukuba, Ibaraki 305-8571, Japan}
\altaffiltext{4}{Department of Physics, School of Science and Technology, Kwansei Gakuin University, 2-1 Gakuen, Sanda, Hyogo 669-1337, Japan}
\altaffiltext{5}{Mizusawa VLBI Observatory, 2-12 Hoshigaoka-cho, Mizusawa, Oshu, Iwate 023-0861, Japan}
\altaffiltext{6}{Institute of Astronomy, The University of Tokyo, 2-21-1 Osawa, Mitaka, Tokyo 181-0015, Japan}

\KeyWords{Galaxy: center -- Galaxy: structure -- ISM: clouds -- ISM: molecules -- Radio lines: ISM} 

\maketitle

\begin{abstract}
 We present survey data of 
 the NH$_3$ $(J,K)=(1,1)$--$(6,6)$ lines,
 simultaneously observed with the Tsukuba 32-m telescope,
 in the main part of the central molecular zone of the Galaxy. 
 The total number of on-source positions was 2655.
 The lowest three transitions were detected with
 ${\rm S/N} > 3$ at 2323 positions (93\% of all the on-source positions).
 Among 2323,
 the S/N of $(J,K) = (4,4),(5,5),$ and $(6,6)$ exceeded 3.0
 at 1426 (54\%), 1150 (43\%), and 1359 (51\%) positions, respectively.
 Simultaneous observations of the lines enabled us
 to accurately derive intensity ratios
 with less systematic errors.
 Boltzmann plots indicate there are two temperature components:
 cold ($\sim 20$ K) and warm ($\sim 100$ K).
 Typical intensity ratios of
 $T_{\rm mb}(2,2)/T_{\rm mb}(1,1)$,
 $T_{\rm mb}(4,4)/T_{\rm mb}(2,2)$,
 $T_{\rm mb}(5,5)/T_{\rm mb}(4,4)$, and
 $T_{\rm mb}(6,6)/T_{\rm mb}(3,3)$ were
 0.71, 0.45, 0.65, and 0.17, respectively.
 These line ratios correspond to diversity of rotational temperature,
 which results from mixing of the two temperature components.
\end{abstract}

\section{Introduction}

The central molecular zone (CMZ) of the Galaxy is 
a highly concentrated region of interstellar molecular gas,
located within about a 200-pc radius from the Galactic Center
(GC; e.g., \cite{morris1996}).
The molecular gas in the CMZ is characterized by the large mass
($5.3\times10^7 M_{\odot}$, \cite{pierce2000}),
large velocity dispersion
(typical line width $\Delta v \approx$ 10--20 km s$^{-1}$),
high number densities ($\sim10^4$ cm$^{-3}$),
and high temperatures.
These characteristics have been shown by various
molecular line observations
(e.g., \cite{jones2013})
including ammonia (NH$_3$) inversion-lines (e.g. \cite{huett1993}).
Meta-stable ($J=K$) inversion-lines of NH$_3$
have been often referred to as an ``interstellar thermometer'',
because intensity ratios between the lines
reflect the kinetic temperature
with tiny influence of variation of the number density
(e.g., \cite{walmsley1983, danby1988}).
The critical densities are not very high
even at levels up to $(J,K)=(15,15)$
whose excitation energy is about 2000 K
[e.g. assuming hot gas ($\sim 300$ K),
 $n_{\rm H_2}\approx10^{3\mathchar`-4}$ cm$^{-3}$,
 \cite{mills2013}].
Therefore, the intensity ratios are the most reliable
and powerful probe of interstellar molecular gas temperature
in very wide temperature range.

Several NH$_3$ observations have been made in the CMZ.
First, long-scan observations along the Galactic longitude were
made by Morris et al. (1983),
revealing that warm temperature
(rotational temperature, $T_{\rm rot}\approx 30$--$60$ K)
is common in the region.
One of the most important reports was made by
H\"{u}ettemeister et al. (1993)
who observed the six lowest meta-stable lines
toward 36 cloud cores in the CMZ,
and showed that there are two temperature components,
cold and warm, in all the clouds.
The cold component ($\sim25$ K) seems to be dense
($\sim10^5$ cm$^{-3}$)
and thermally coupled with dust
($T_{\rm dust}\approx$14--20 K, \cite{pierce2000}),
while the warm component ($> 100$ K) seems to be less dense
($\sim10^4$ cm$^{-3}$).
Large-scale mapping observations were carried out
recently by Nagayama et al. (2007, 2009) and Purcell et al. (2012).
Nagayama et al. (2007, 2009)
argued that the envelopes of molecular clouds are hotter than
the insides
and the ortho-para ratio (OPR) is high in the CMZ ($\sim$1.5--3.5).
Extremely hot NH$_3$ excited up to $(J,K)=(18,18)$ was detected
toward the core region of Sgr B2 as absorption lines (\cite{wilson2006}).
Mills and Morris (2013) detected NH$_3$ $(J,K)=(8,8)$--$(15,15)$ lines
toward 15 cloud centers,
revealing that hot (over 200--300 K) molecular gas exists
in various places in the CMZ. 
Similar warm (or hot) gas was also discovered by other line observations,
such as H$_2$ emissions (\cite{rod2001}),
H$_3^+$ absorption lines (\cite{goto2008}),
and H$_3$O$^+$ absorption lines (\cite{lis2010}).
Their distributions, however, remain unrevealed.

Warm or hot gas is likely to be ubiquitous in the CMZ.
Heating of molecular gas by dust,
however, does not work in this region,
because the temperature of the warm component is significantly higher
than that of dust (e.g. \cite{pierce2000}). 
Several other candidates for the heating mechanism have been proposed; mainly,
(1) X-ray heating (e.g., \cite{nagayama2007}),
(2) cosmic ray heating (e.g., \cite{guesten1981,ao2013}),
(3) dissipation of supersonic turbulence produced
by shock phenomena (e.g., \cite{wilson1982, flower1995, riquelme2013}).
The mechanisms (1) and (2) appear insufficient to heat
the ubiquitous hot component with moderately intense X-ray
(flux density, $F_X\approx2.0\times10^{-3}$ erg cm$^{-2}$ s$^{-1}$,
 \cite{koyama2007, ao2013})
and cosmic-ray (ionization rate, $\zeta\approx$ a few 10$^{-14}$ s$^{-1}$, \cite{yusef2007})
in the GC.
On the other hand, several studies estimated that
heating via shock phenomena can be attributed to the heat mechanism
of molecular gas
(e.g. \cite{wilson1982, flower1995, mills2013}).
Sources of shocks have been observationally proposed
in this region so far:
(1) supernova or hypernova explosions (e.g. \cite{tanaka2009}),
(2) collisions between molecular clouds (\cite{hasegawa1994, menten2009, mills2013, tsuboi2015}).
Collisions of molecular gas may be partly due to the barred potential of the Galaxy (\cite{binney1991}),
or magnetic fields such as Parker instability (\cite{fukui2006}).
To study how these shocks contribute to the heating,
investigation of the distribution and morphology of
warm/hot molecular gas is intrinsically important.

We mapped NH$_3$ meta-stable inversion-lines $(J,K)=(1,1)$--$(6,6)$ simultaneously in the CMZ
with the Tsukuba 32-m telescope owned by the Geospatial Information Authority of Japan (GSI).
In this paper, we describe observations, data analysis process, and present measured data.
Detailed analysis of the data and discussion of mechanisms
heating molecular clouds in the CMZ will be made in the forthcoming paper.

\section{Observations and data reduction}\label{sec:obs}

Observations of the NH$_3$ $(J,K)=(1,1)$--$(6,6)$
inversion-transition lines were made over 50 days from 2009 to 2012
with the Tsukuba 32-m telescope.
The basic performance of the telescope is
shown in table \ref{tab:tele}.
These lines were observed simultaneously.
The rest frequencies and excitation energies are
summarized in table \ref{tab:nh3}.
A radio recombination line (RRL) H64$\alpha$ (24.50991 GHz) was also observed
at the same time.
The half power beam width (HPBW) of the telescope was $\timeform{93''}\pm\timeform{6''}$
at $24.0$ GHz,
corresponding to 3.6 pc at the distance to the GC, 8.05 kpc (\cite{honma2012}).
The main beam efficiency, $\eta_{\rm mb}$, of the antenna
was measured by
observing Jupiter whose brightness temperature was adopted to be
$T_{\rm b}=138\pm7$ K at $1.3$ cm (23 GHz, \cite{depater2005}).
The efficiency depended on the elevation angle (EL) of the antenna as
\begin{equation}
\eta_{\rm mb}({\rm EL})
= 0.3337 +7.435 \times 10^{-3} {\rm EL}
- 7.159\times 10^{-5} {\rm EL}^2 
+4.355\times10^{-7} {\rm EL}^3
\end{equation}
at 24.0 GHz for ${\rm EL}=$\timeform{5D}--\timeform{80D}
with the maximum value of 
$\eta_{\rm mb}=0.49\pm0.02$ at ${\rm EL}=\timeform{38D.4}$.

\begin{table}
 \caption{Basic performance of the Tsukuba 32-m antenna of GSI.}\label{tab:tele}
 \begin{center}
  \begin{tabular}{rl}
   \hline \hline
   Location & \timeform{36D06'11''} N, \timeform{140D05'19''} E\\
   Altitude & 44.6 m\\
   Antenna mount & Cassegrain, altazimuth\\
   Aperture diameter & 32 m\\
   Frequency bands & 19.5--25.1 GHz (K-band)\\
   Aperture efficiency & $\leq0.41\pm0.02$ (at EL$\approx\timeform{38D}$)\\
   Main beam efficiency & $\leq0.49\pm0.02$ (at EL$\approx\timeform{38D}$)\\
   Beam size & \timeform{93''}$\pm$\timeform{6''} at 24.0 GHz\\
   Polarization & right-hand circular polarization (--2009)\\
   & right- and left-hand circular polarizations (2010--)\\
   Backend bandwidth & 1.0 GHz $\times$ 2\\
   Spectral resolution & 61 kHz\\
   \hline
  \end{tabular}
 \end{center}
\end{table}

\begin{table}
 \caption{Properties of NH$_3$ rotational inversion-transitions and H64$\alpha$.}\label{tab:nh3}
 \begin{center}
  \begin{tabular}{ccccc}
   \hline \hline
   Transition $(J,K)$ & Frequency [GHz] & $E_{\rm u}/k_{\rm B}$ [K] & $a_{\rm in}/a_{\rm out}$\footnotemark[$*$] & $v_{\rm in}/v_{\rm out}$ [km s$^{-1}$]\footnotemark[$\dagger$]\\ \hline
   $(1,1)$ & 23.694496 & 22.1 & 0.2778/0.2222 & 7.7/19.4\\
   $(2,2)$ & 23.722633 & 63.3  & 0.0651/0.0628 & 16.3/26.1 \\
   $(3,3)$ & 23.870129 & 122.4 & 0.0300/0.0296 & 21.1/29.2 \\
   $(4,4)$ & 24.139416 & 199.4 & 0.0174/0.0173 & 23.7/30.9 \\
   $(5,5)$ & 24.532989 & 294.2 & 0.0114/0.0114 & 25.4/31.9 \\
   $(6,6)$ & 25.056025 & 406.9 & 0.0081/0.0081 & 26.3/32.1 \\ 
   H64$\alpha$ & 24.50990\footnotemark[$\ddag$]  &  &  &  \\ \hline
  \multicolumn{5}{@{}l@{}}{\hbox to 0pt{\parbox{120mm}{\footnotesize
  \par\noindent
  \footnotemark[$*$] $a_{\rm in}$ and $a_{\rm out}$ are transition probabilities of the satellite lines
  (intensity relative to the main line in optically thin limit),
  which were theoretically calculated by a formula shown in the bottom of
  page 87 in Kukolich (1967).
  \par\noindent
  \footnotemark[$\dagger$] $v_{\rm in}$ and $v_{\rm out}$ are frequency offsets of the satellite lines
  from the main line in velocity unit, which were quoted from Simmons and Gordy (1948).
  \par\noindent
  \footnotemark[$\ddag$] Gordon and Sorochenko (2002)
  }\hss}}
  \end{tabular}
 \end{center}
\end{table}

The receiver front-end was a feed-horn
with an ortho-mode transducer for circular polarizations,
succeeded by a HEMT amplifier for each output cooled to 11 K.
Right-hand circular polarization observations were started in 2009,
while components for left-hand circular polarization were
installed in 2010,
and then dual polarizations were observed simultaneously to reduce noise level.
We used two FX-type 8-bit digital spectrometers
for each polarization.
The bandwidth and spectral resolution of the spectrometers
were 1 GHz and 61 kHz,
corresponding to 12491 km s$^{-1}$ and 0.76 km s$^{-1}$
at 24 GHz, respectively.
The spectrometers covered
23.6--24.6 GHz for $(J,K)=(1,1)$ to $(5,5)$
and 24.6--25.1 GHz for $(6,6)$, respectively.

Intensity calibration was carried out by the chopper-wheel method,
yielding antenna temperature $T_{\rm A}^*$,
corrected for both atmospheric and antenna ohmic losses
(\cite{ulich1976, kutner1981}).
The main beam brightness temperature,
$T_{\rm mb}(\equiv T_{\rm A}^*/\eta_{\rm mb})$, was converted
from $T_{\rm A}^*$ using  the main beam efficiency $\eta_{\rm mb}$,
at each observing elevation and frequency.
Daily variation of the intensity scale was
checked by observing Sgr B2 once a day.

\begin{figure}[htbp]
 \begin{center}
  \FigureFile(160mm,20mm){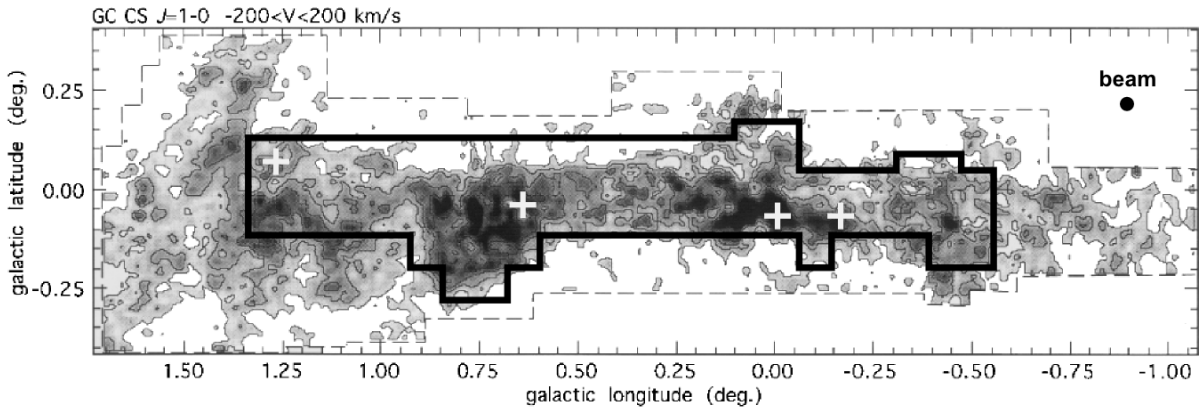}
 \end{center}
 \caption{Survey area surrounded by a thick line overlaid with
          the CS (1--0) integrated intensity map
          (grey scale; \cite{tsuboi1999}).
          The filled circle at the upper right represents
          the beam size of the 32-m telescope (HPBW $=\timeform{93''}$).
          The four grey crosses indicate the positions of spectrum drawn
          in figure \ref{fig:spectra}.}
          \label{fig:csmap}
\end{figure}

During the observations,
the typical system noise temperatures, $T_{\rm sys}$,
were 75--200 K.
Figure \ref{fig:csmap} shows the observed area.
A filled circle depicted at the upper right represents
the main beam size of the 32-m telescope (\timeform{93''}).
The observations were made with the position-switching method.
Reference position was $(L,B)=(\timeform{1D.0}, \timeform{-0D.5})$ or $(\timeform{-0D.5}, \timeform{-0D.5})$.
On-source positions were
on a grid with an origin of $(L,B)=(\timeform{0D},\timeform{0D})$
and at intervals of $\timeform{50''}$,
which covered the sky with slight undersampling.
The total number of on-source positions was 2655.
On-source integration time was from 30 seconds to 4 minutes,
resulting in noise levels of 0.1--1.0 K for the $(1,1)$ line.
To focus on spatial structure larger than the beam size,
we applied spatial smoothing, described later.
Pointing errors were corrected every hour
by observing an H$_2$O maser of VX Sgr at 22.235 GHz.
The typical pointing accuracy was $\sim\timeform{25''}$.

\begin{figure}[htbp]
 \begin{center}
  \FigureFile(160mm,20mm){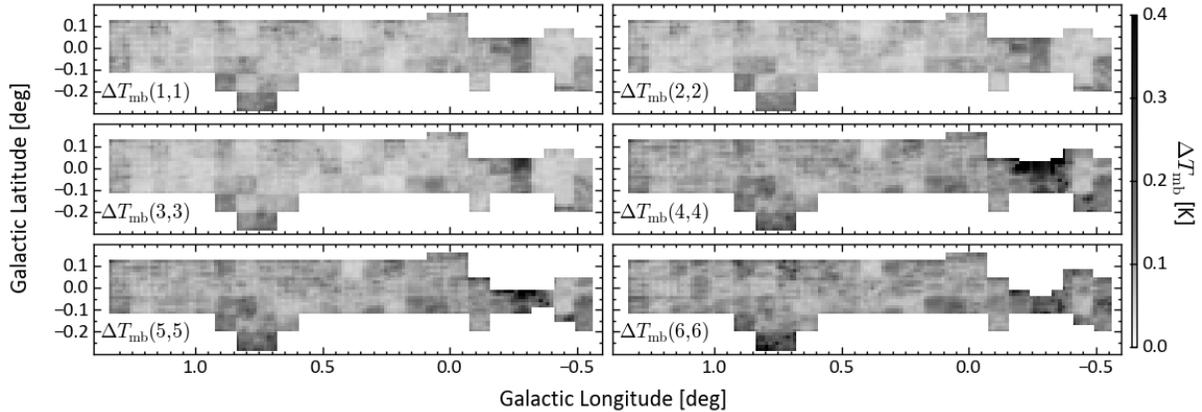}
  \label{fig:RMSmap}
 \end{center}
 \caption{RMS maps of the six NH$_3$ lines.}
\end{figure}

All data were reduced using the software package NEWSTAR
developed by the Nobeyama Radio Observatory (NRO).
We separately flagged the six NH$_3$ line data with bad baselines.
In the $(5,5)$ and $(6,6)$,
several positions lost all the data due to flagging and
are dropped from the maps.
The baselines of the spectra were fitted with a linear function
and subtracted.
The spectra were binned in the 2 km s$^{-1}$ velocity width,
sufficiently narrow compared to the typical velocity width (10--20 km s$^{-1}$).
Then, the data cube was spatially smoothed with
a Gaussian weighting function of the full width half maximum
(FWHM) of $\timeform{100''}$,
resulting in the effective spatial resolution of
$\theta_{\rm Tsu}=\timeform{137''} (5.3\ {\rm pc})$.
After the smoothing,
the noise level was reduced to
typically $\Delta T_{\rm mb}\approx0.09$ K
as shown in RMS maps of figure \ref{fig:RMSmap}.

\begin{table}
 \caption{Sammary of Tsukuba 32-m NH$_3$ survey.}\label{tab:survey}
 \begin{center}
  \begin{tabular}{rl}
   \hline \hline 
   Observed lines & NH$_3$ $(J,K)=(1,1)$--$(6,6)$ (simultaneously)\\
   Typical system noise temperature & 75--200 K (in $T_{\rm A}^*$ scale)\\
   Interval of grid points & $\timeform{50''}$ ($2.0$ pc at 8.05 kpc)\\
   Total number of on-source position & 2655\\
   Effective spatial resolution & $\timeform{137''}$ (5.3 pc at 8.05 kpc)\\
   Typical RMS noise level & $\sim 0.09$ K (smoothed data in $T_{\rm mb}$ scale)\\
   \hline
  \end{tabular}
 \end{center}
\end{table}

\section{Results}

\subsection{Spectra}\label{subsec:integmaps}

All signal-to-noise ratio (S/N)
of the $(J,K) = (1,1)$, $(2,2)$, and $(3,3)$ lines
simultaneously exceeded 3
at 2323 positions (87 \% out of all observed positions).
At these positions, S/N of the $(J,K)=(4,4)$, $(5,5)$,
and $(6,6)$ lines
exceeded 3 at 1426 (54\%), 1150 (43\%),
and 1359 (51\%) positions, respectively.
The intensity of the $(3,3)$ line was the strongest
at almost all observed positions and that of $(1,1)$
the second strongest.
This trend was also observed in previous research efforts
(e.g. \cite{morris1983, nagayama2009}).

\begin{figure}[htbp]
 \begin{center}
  \FigureFile(160mm,60mm){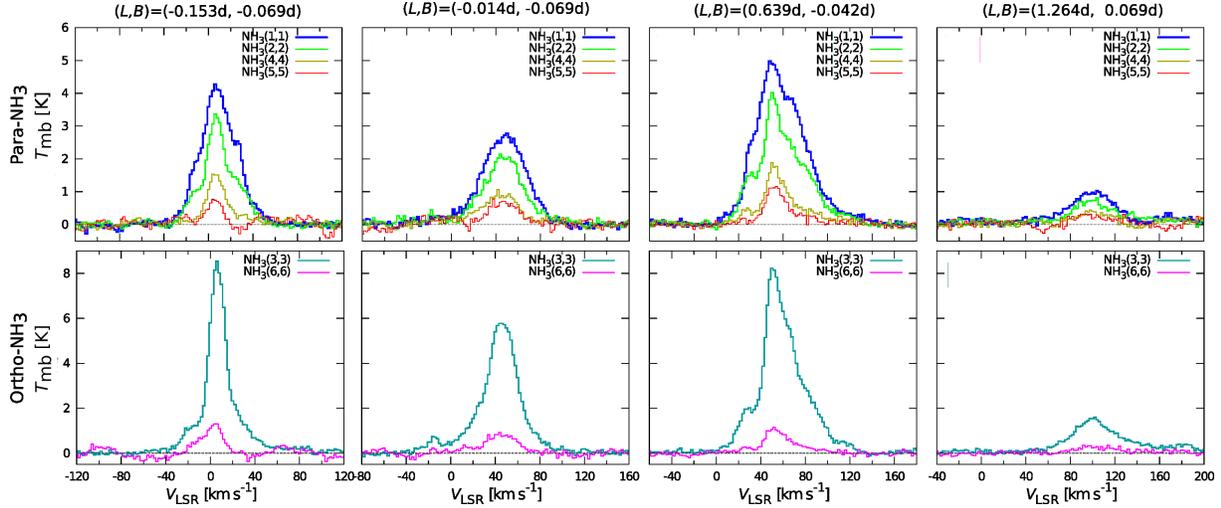}
  \end{center}
 \caption{
 Examples of NH$_3$ spectra at four positions.
 Upper panels are para-NH$_3$ lines and
 lower panels are ortho-NH$_3$ lines.
 {\it Left}: $(L,B)=(\timeform{-0D.153},\timeform{-0D.069})$
 is in the 20-km s$^{-1}$ cloud.
 {\it Middle left}:
 $(L,B)=(\timeform{-0D.014},\timeform{-0D.069})$
 is in the 50-km s$^{-1}$ cloud.
 {\it Middle right}:
 $(L,B)=(\timeform{0D.639},\timeform{-0D.042})$
 is in the Sgr-B2 region.
 {\it Right}:
 $(L,B)=(\timeform{1D.264},\timeform{0D.069})$ is
 in the cloud of the $\timeform{1D.3}$ region.
 }\label{fig:spectra}
\end{figure}

The NH$_3$ rotational-inversion lines has
five groups of hyperfine lines in a narrow velocity range
($<5.5$ MHz $\approx 70$ km s$^{-1}$)
[table \ref{tab:nh3}; Kukolich (1967)].
They overlap with each other, however,
due to their large velocity widths ($\approx 20$ km s$^{-1}$) toward the CMZ
[the features of the hyperfine lines and the overlapping effect in the CMZ
 is well expounded by earlier studies; e.g. McGary and Ho (2002)].
Figure \ref{fig:spectra} shows sample spectra of
NH$_3$ lines at four prominent positions
``the 20--km s$^{-1}$ cloud'' (GCM$-$0.13$-$0.08),
``the 50--km s$^{-1}$ cloud'' (GCM$-$0.02$-$0.07),
the Sgr-B2 cloud, and the $L=\timeform{1D.3}$ region.

\subsection{Distributions of Line Intensities}\label{sec:dist}

\begin{figure}[htbp]
 \begin{center}
  \FigureFile(160mm,20mm){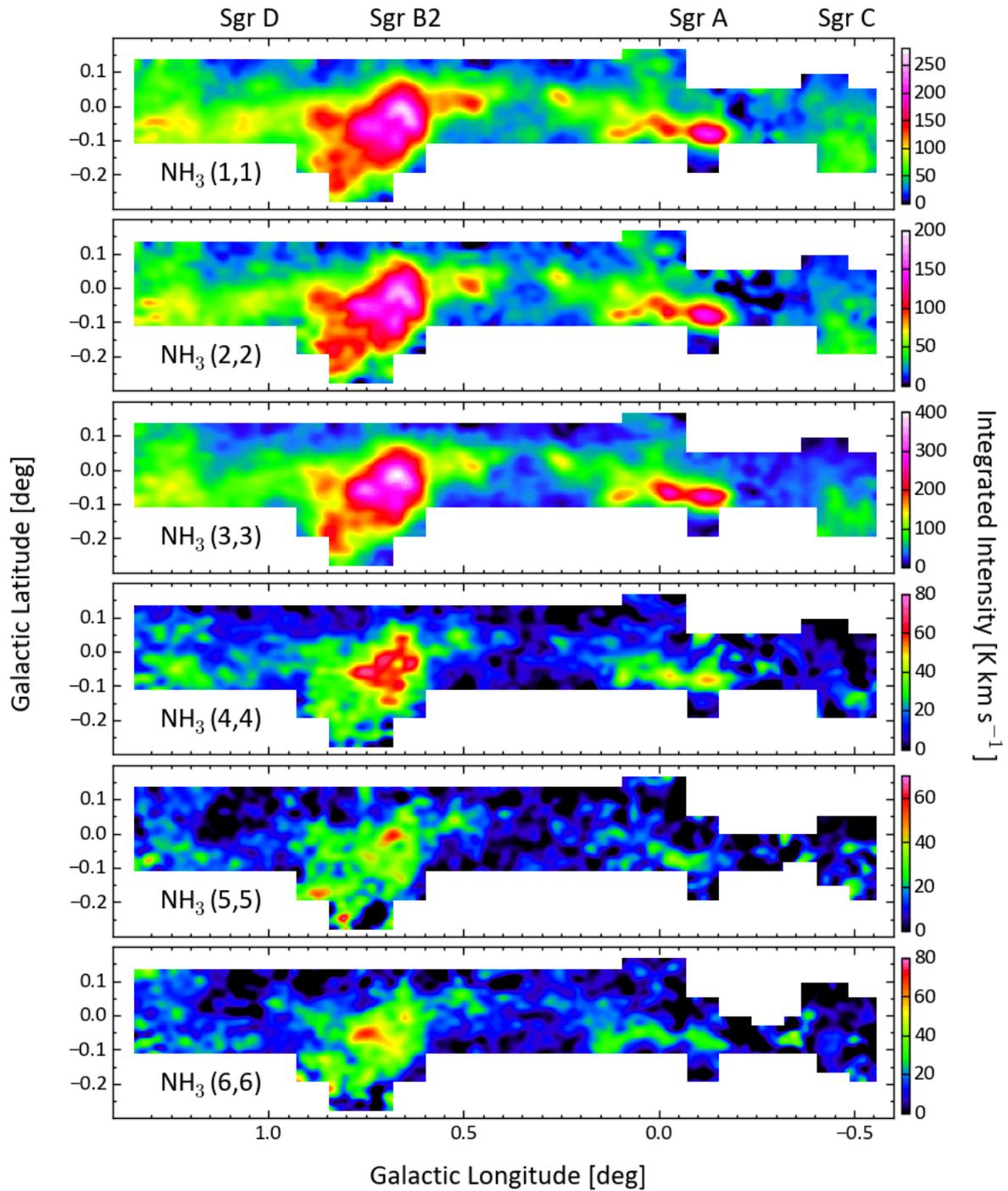}
 \end{center}
 \caption{
 Integrated intensity maps of NH$_3$ $(1,1)$--$(6,6)$.
 The integrated velocity range is
 $V_{\rm LSR}=-200$ to $200$ km s$^{-1}$.
 These maps were spatially smoothed with a Gaussian function,
 resulting in the effective resolution of $\timeform{137''}$.}\label{fig:integ}
\end{figure}

Figure \ref{fig:integ} shows
the spatial distributions of the integrated intensity,
$I\equiv\int T_{\rm mb}dv$, of NH$_3$ $(J,K)=(1,1)$--$(6,6)$.
Several molecular cloud complexes
such as Sgr A ($L\approx\timeform{0D}$), Sgr B2 ($L\approx\timeform{0D.7}$),
Sgr C ($L\approx\timeform{-0D.5}$), and the $L = \timeform{1D.3}$ region
can be clearly seen in all the maps.
The distributions of $(J,K)=(1,1)$--$(3,3)$ lines
are quite similar each other.
Maps of higher transitions, $(J,K)=(4,4)$--$(6,6)$, also resemble
but are more clumpy than the $(1,1)$--$(3,3)$ lines.

RRL H64$\alpha$
was significantly detected toward only two positions,
the Sgr-B2 core region and the Sgr-B1 region
[ $(L,B)=(\timeform{0D.667}, \timeform{-0D.028})$ and
 $(\timeform{0D.528}, \timeform{-0D.056})$,
 with the typical $\Delta T_{\rm mb}\sim 0.13$ K].
The result is consistent with earlier research (e.g. Jones et al. 2012, 2013).
Although there are several other RRLs situated close to the six NH$_3$ lines (e.g. H82$\beta$),
 they were not detected in our observed data.
Therefore, the NH$_3$ data are not influenced by RRLs outside the two positions
(i.e. the integrated intensity maps are not polluted
 by other than the six NH$_3$ lines
 with the exception of the Sgr-B2 core region).

\begin{figure}[htbp]
 \begin{center}
  \FigureFile(160mm,20mm){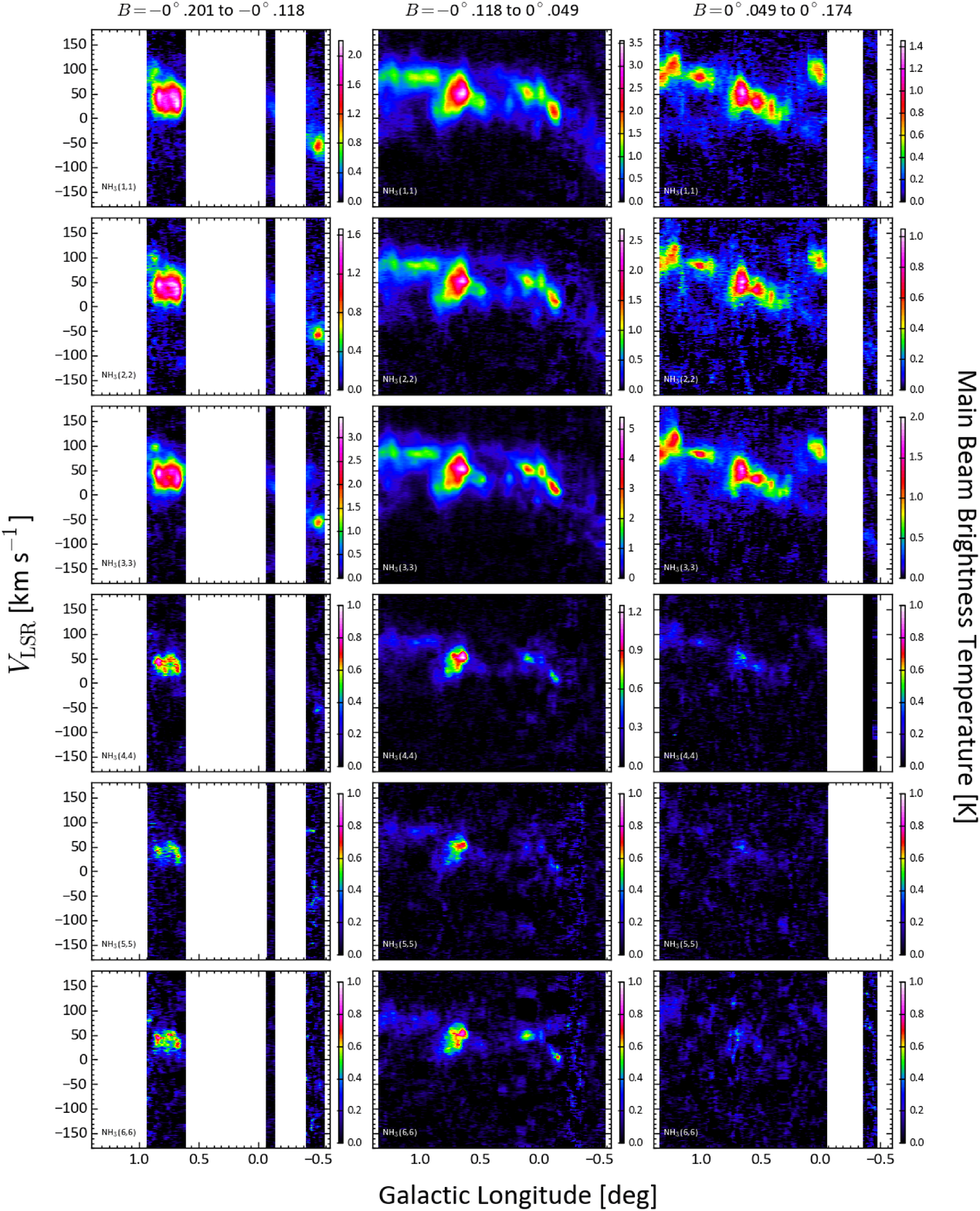}
 \end{center}
  \caption{The longitude-velocity diagrams of the main beam brightness temperature $T_{\rm mb}$ of
           NH$_3$ $(J,K)=(1,1)$--$(6,6)$ averaged over latitudes of
           $B=\timeform{-0D.201}$--$\timeform{-0D.118}$ (left),
           $B=\timeform{-0D.118}$--$\timeform{+0D.049}$ (center) and
           $B=\timeform{+0D.049}$--$\timeform{+0D.174}$ (right).
           }\label{fig:LVmap}
\end{figure}

Figure \ref{fig:LVmap} shows the longitude-velocity diagrams
with the latitude-range divided into three regions:
averaged over
$B=\timeform{-0D.285}$ to $\timeform{-0D.118}$ (left column),
$B=\timeform{-0D.118}$ to $\timeform{0D.049}$ (center column)
and $B=\timeform{0D.049}$ to $\timeform{0D.176}$ (right column),
respectively.
The pixel size is $\timeform{10''}$ $\times$ 2 km s$^{-1}$.
No absorption due to foreground gas in the Galactic disk region
was found in our NH$_3$ data,
the same as in previous research (e.g. \cite{nagayama2009}).

\subsection{Comparison with Other NH$_3$ Data}\label{sec:mopra}

To verify the $T_{\rm mb}$ scaling,
we compared our data with a previous survey of NH$_3$ $(J,K)=(1,1)$ and $(2,2)$
around the CMZ carried out with the Mopra 22-m telescope
(\cite{walsh2011, purcell2012},
the spatial resolution $\theta_{\rm Mop}=\timeform{2'}$,
data are available online).
We summed the Mopra data over each 2 km s$^{-1}$ velocity bin
and smoothed them with a Gaussian function, which has the FWHM of
$\sqrt{(\theta_{\rm Tsu}^2-\theta_{\rm Mop}^2)}\approx\timeform{65''}$
to match the spatial resolution with that of our data.

\begin{figure}[htbp]
 \begin{center}
  \FigureFile(90mm,70mm){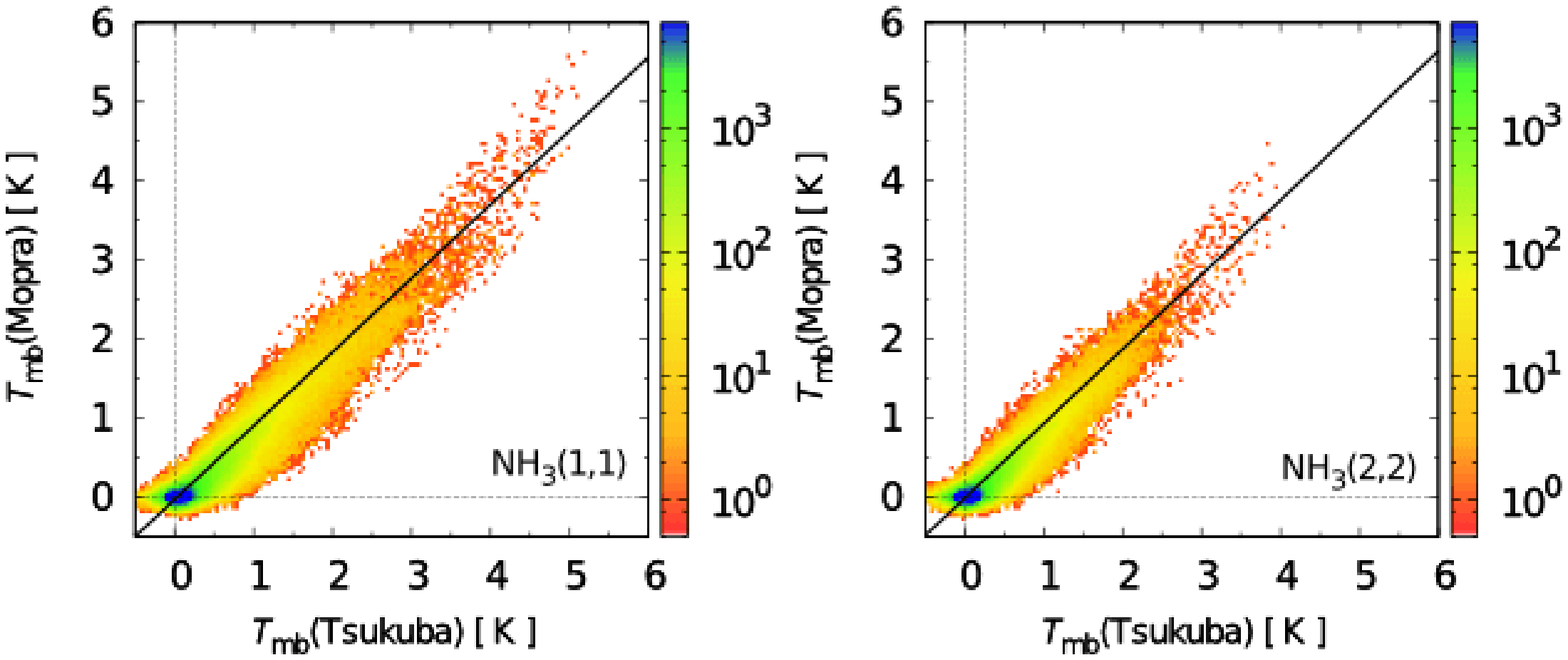}
 \end{center}
 \caption{
 Intensity correlation of the two NH$_3$ inversion-lines
 between the Tsukuba 32-m telescope
 and the Mopra 22-m telescope (\cite{purcell2012}).
 The pixel size is 0.05 K $\times$ 0.05 K.
 Solid lines are the result of the linear fitting;
 $T_{\rm mb}({\rm Mopra})=0.938\times T_{\rm mb}({\rm Tsukuba})-0.033$ for NH$_3$ $(1,1)$ and
 $T_{\rm mb}({\rm Mopra})=0.944\times T_{\rm mb}({\rm Tsukuba})-0.019$ for NH$_3$ $(2,2)$.
 Details are explained in subsection \ref{sec:mopra}.
 }\label{fig:mopra}
\end{figure}

The results are shown in figure \ref{fig:mopra}
as a plot of the intensity correlation
between our NH$_3$ data and the Mopra data in the $T_{\rm mb}$ scale.
The correlation is well fitted with a linear function,
$T_{\rm mb}({\rm Mopra})=0.938\times T_{\rm mb}({\rm Tsukuba})-0.033$ for $(1,1)$ and
$T_{\rm mb}({\rm Mopra})=0.944\times T_{\rm mb}({\rm Tsukuba})-0.019$ for $(2,2)$, respectively, where
the fitting is weighted according to the S/N of each pixel, that is,
using the weight,
\begin{equation}
 w=\left[ \left( \frac{\sigma_{\rm Tsu}}{T_{\rm mb}({\rm Tsukuba})} \right)^2
         +\left( \frac{\sigma_{\rm Mop}}{T_{\rm mb}({\rm Mopra})} \right)^2 \right]^{-0.5},
\end{equation}
where $\sigma_{\rm Tsu}$ is the noise level shown in figure \ref{fig:RMSmap},
and $\sigma_{\rm Mop}$ is the typical noise level of the Mopra data, $0.08$ K.
We confirm that the difference of the $T_{\rm mb}$ scale of our data from Mopra data is less than $10\%$.

\subsection{Boltzmann Plots}\label{sec:blp}

We can grab the excitation state with Boltzmann plots
(i.e. rotational diagram, \cite{goldsmith1999}) from multi-line observations.
In a Boltzmann plot,
column densities of each $(J,K)$ normalized by the statistical weight
versus the excitation energy are plotted.
If a single temperature gas predominates the observed NH$_3$ gas,
the normalized NH$_3$ column densities lie on
a single straight slope of inverse proportion to the rotational temperature $T_{\rm rot}$.

\begin{figure}[htbp]
 \begin{center}
  \FigureFile(160mm,80mm){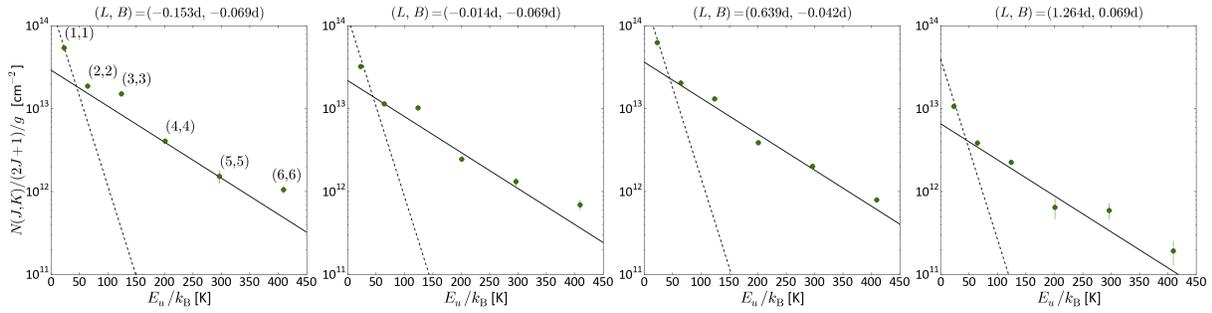}
 \end{center}
 \caption{
 Boltzmann plots of the intensity peak voxel
 at four points of figure \ref{fig:spectra}.
 The velocity ranges are
 3--9, 39--45, 45--51, and 87--93 km s$^{-1}$, respectively.
 Lines with a constant rotation temperature
 for cold ($20$ K, dotted) and warm ($100$ K, solid) are also shown as a guide.
 Column densities of LTE lines both cold and warm are
 $4\times10^{14}$ cm$^{-2}$, $3\times10^{14}$ cm$^{-2}$,
 $5\times10^{14}$ cm$^{-2}$ and $9\times10^{13}$ cm$^{-2}$, respectively (from left to right).
 }\label{fig:bolt}
\end{figure}

Figure \ref{fig:bolt} shows four Boltzmann plots of our NH$_3$ data
for the peak voxel at the same four positions as figure \ref{fig:spectra}.
The normalized column densities are calculated by the formula under assumptions
of the local thermodynamic equilibrium (LTE)
and thin optically depth
\begin{equation} \label{eq:Nbolt}
\frac{N(J,K)}{g(2J+1)}=\frac{3 k_{\rm B}}{4 \pi^3 \nu(J,K) |\mu(J,K)|^2 g(2J+1)} T_{\rm mb}(J,K) \Delta \nu,
\end{equation}
where $g$ is statistical weight ($4$ for $K=3,6$ and 2 for the others),
$k_{\rm B}$ the Boltzmann constant,
$\nu(J,K)$ the line frequency,
$|\mu(J,K)|^2=\mu^2 K^2/\left[J(J+1)\right]$,
$\mu$ the permanent electric dipole moment of NH$_3$, 1.468 debye,
$T_{\rm mb} \Delta \nu$ the integrated intensity of a voxel.
Here we used a data cube of the velocity width $\Delta \nu$ smoothed to 6 km s$^{-1}$
(i.e., the voxel size was
$\timeform{50''}\times\timeform{50''}\times6$ km s$^{-1}$),
which allows us to investigate physical conditions of each velocity component.
The typical RMS noise level is about $T_{\rm mb}=0.07$ K.
Because there is satellite emission of the hyperfine splitting out of the voxel velocity range,
$N(1,1)$ is now underestimated by factor 0.5--1.0.

The data points in figure \ref{fig:bolt} shows that
populations in higher levels are warm, about 100 K;
we draw a line of 100 K as a guide.
Note that the ortho-NH$_3$ points are slightly above the line
because of OPR of $>1$.
The points of $(1,1)$ are also above the 100--K line,
indicating that there are a cold component.
We draw another LTE line of 20 K
whose column density is set to that of the 100--K line in each plot,
which become close to the $(1,1)$ points.
H\"{u}ettemeister et al. (1993) reported that
the two temperature components are needed to elucidate the distribution of
the NH$_3$ toward the cloud cores observed by them.
The two temperature components are universal in our data,
not only toward the cloud cores,
which is shown in terms of intensity ratios in subsection \ref{sec:iir}.
Further detailed analysis of the two temperature components
including derivation of the physical state using model fitting is shown in the subsequent paper.

\subsection{Intensity Ratio Distribution in Cube Data}\label{sec:iir}

We show distributions of the intensity ratios
between two of lines: $R_{21}$, $R_{42}$, and $R_{54}$ of para-NH$_3$,
and $R_{63}$ of ortho-NH$_3$,
where $R_{ul}\equiv T_{\rm mb}(u,u)/T_{\rm mb}(l,l)$
at each voxel same as subsection \ref{sec:blp}
(i.e., $\timeform{50''}\times\timeform{50''}\times6$ km s$^{-1}$).
The intensity ratios are important as
the simplest indicator of molecular gas temperature,
doing not involve complex derivation of physical parameters.
The ratios in this subsection were calculated only at voxels where emission of the lower transition
was detected over $5\sigma$.

\begin{figure}[htbp]
 \begin{center}
  \FigureFile(150mm,70mm){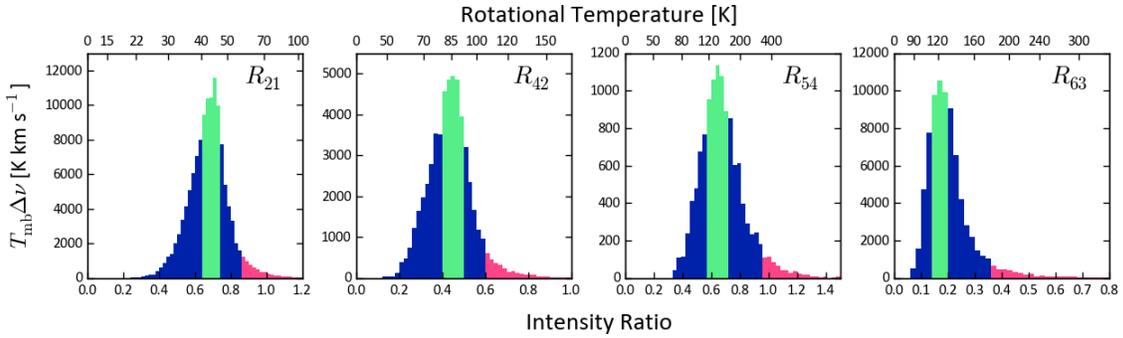}
 \end{center}
 \caption{
 Histograms of the intensity ratios,
 $R_{ul}\equiv T_{\rm mb}(u,u)/T_{\rm mb}(l,l)$,
 weighted by the integrated intensities of the lower transitions,
 that is,
 the ordinate is the summation of the integrated intensities of
 the lower transition, $T_{\rm mb}(l,l) \Delta \nu$.
 The width of the bins is $\Delta R=0.02$.
 In each histogram,
 the predominant ratio range (PR) and
 the higher ratio range (HR) are indicated by
 light green and pink, respectively.
 }\label{fig:ratio_histo}
\end{figure}

Figure \ref{fig:ratio_histo} shows
the histograms of the intensity ratios
where the ordinate is the summation of
the integrated intensity of the lower transition, $T_{\rm mb}(l,l) \Delta \nu$.
When the lines are optically thin,
intensity ratios reflect the physical condition of gas, especially the kinetic temperature.
We take the value of the peak in the histograms
as a typical value (TV) of each intensity ratio.
Here, we define two ranges in the histograms to characterize
the distributions of the intensity ratios:
predominant ratio range (PR) and higher ratio range (HR).
PR is the range collecting the largest $T_{\rm mb}(l,l)$ bins
so that 40\% of the total integrated intensity
of the lower line is included in.
HR is the range collecting the highest ratio bins including 5\%
of the intensity.
PR and HR are filled with green and pink in figure \ref{fig:ratio_histo}.
These values are summarized in table \ref{tab:ratio}.
The TV of $R_{21}$ (0.71) is very close to
the mean value of $R_{21}$ reported in earlier studies
($\sim 0.70$, \cite{nagayama2009}).

\begin{table}
 \caption{Summary of intensity ratios and corresponding rotational temperatures}\label{tab:ratio}
 \begin{center}
  \begin{tabular}{cccc}
   \hline \hline
   & Typical ratio ($T_{\rm rot}$) & Predominant ratio range ($T_{\rm rot}$) & Higher ratio range ($T_{\rm rot}$)\\
   \hline
   $R_{21}$ & 0.71 (36 K)  & 0.64--0.74 (33--38 K) & $>0.86$ ($>43$ K)\\
   $R_{42}$ & 0.45 (86 K)  & 0.40--0.50 (80--92 K) & $>0.60$ ($>105$ K)\\
   $R_{54}$ & 0.65 (138 K) & 0.57--0.72 (116--163 K) & $>0.96$ ($>320$ K)\\
   $R_{63}$ & 0.17 (111 K) & 0.14--0.20 (103--119 K) & $>0.36$ ($>157$ K)\\
   \hline
   \multicolumn{4}{@{}l@{}}{\hbox to 0pt{\parbox{180mm}{\footnotesize
   \par\noindent
  }\hss}}
  \end{tabular}
 \end{center}
\end{table}

From the intensity ratios,
we derived the corresponding rotational temperatures, $T_{ul}$,
using the Boltzmann equation
under assumptions of thin optical depths, same beam filling factors, and LTE,
\begin{equation}
\frac{T_{\rm mb}(u,u)}{T_{\rm mb}(l,l)}
 = \frac{\nu(u,u)S(u,u)}{\nu(l,l)S(l,l)} \exp \left( -\frac{\Delta E_{ul}}{k_{\rm B }T_{ul}} \right)
\end{equation}
where $S(J,K)=(2J+1)K^2/\left[J(J+1)\right]$,
and $\Delta E_{ul}$ is the energy difference between level $u$ and $l$.
Table \ref{tab:ratio} shows the resultant rotational temperatures.
The differences among the rotational temperatures are
mainly due to co-existence of two temperature components shown
in subsection \ref{sec:blp}.
The cold component has substantial contributions
to the emission of $T_{\rm mb}(1,1)$,
and little contributions to the emission
of the other higher transitions over $T_{\rm mb}(3,3)$.
Therefore, the rotational temperature $T_{21}$ indicates
the temperature of the cold gas component,
and $T_{54}$ and $T_{63}$ indicate that of the warm (or hot).
$T_{42}$ is contributed from both cold and warm gas components.

\begin{figure}[htbp]
 \begin{center}
  \FigureFile(160mm,210mm){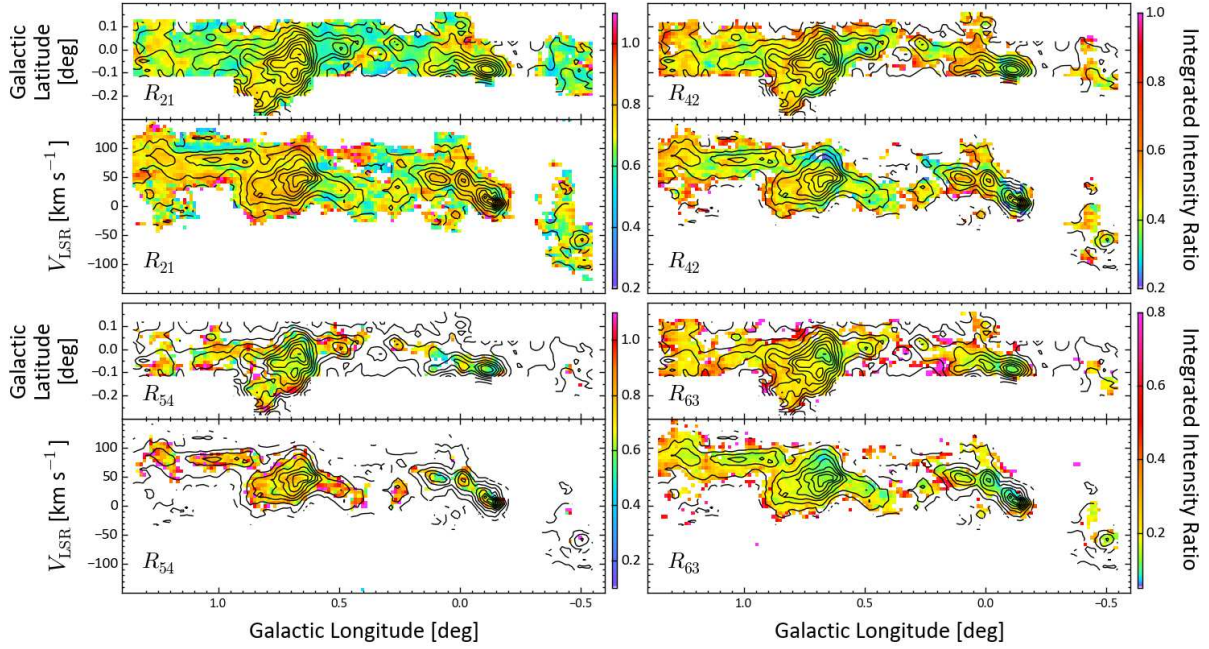}
 \end{center}
 \caption{
 Distributions of the intensity ratio,
 $\overline{R_{ul}}$ (color),
 overlaid on the distributions of the intensity
 of the $(J,K)=(2,2)$ (contours).
 LB-maps and LV-diagrams are shown alongside.
 The contour interval is 18 K km s$^{-1}$ for LB-maps
 and 0.2 K for LV-diagram.
 }\label{fig:average_ratio_map}
\end{figure}

Figure \ref{fig:average_ratio_map} shows
the distributions of the intensity ratio
in the longitude-latitude (LB) map and
the longitude-velocity (LV) diagram.
To project 3-D voxel data to 2-D maps or diagrams,
we calculated the integrated intensity ratio $\overline{R_{ul}}$
along the velocity axis or the galactic latitude axis.
They are expressed as
\begin{eqnarray}
 \overline{R_{ul}(L,B)} = \sum_{v} T_{\rm mb}(u,u)(L,B,v) \Delta v \bigg/  \sum_{v} T_{\rm mb}(l,l)(L,B,v) \Delta v,
\end{eqnarray}
and for LV diagrams,
\begin{eqnarray}
 \overline{R_{ul}(L,v)} = \sum_{B} T_{\rm mb}(u,u)(L,B,v) \Delta v \bigg/  \sum_{B} T_{\rm mb}(l,l)(L,B,v) \Delta v,
\end{eqnarray}
where $\Delta v$ and $\Delta B$ are the width of the 3-D voxel along
the velocity axis and the galactic-latitude axis, respectively.
The LB maps are overlaid with contours of
the integrated intensities of $T_{\rm mb}(2,2)$
and the LV diagrams are overlaid with
contours of $T_{\rm mb}(2,2)$ averaged along $B$.
The inversion-lines of NH$_3$ have satellite lines
with different apparent velocities (frequencies)
from the main lines,
and they would affect the intensity ratios at higher and lower frequencies
(the edges of the spectrum) in the average calculation.
For highly excited NH$_3$ transition lines, however,
the effect can be ignored in the calculations,
because the intensity of satellite lines
is quite weak compared with that of the main line
(see table \ref{tab:nh3}).
The most noticeable feature
in the LB maps of figure \ref{fig:average_ratio_map}
is the distribution of $\overline{R_{63}}$,
namely the intensity ratio is
higher at outer parts of giant molecular clouds (GMCs)
than at inner parts
and a temperature gradient of warm component gas can be seen.
This trend is also seen in the LV-diagram:
a higher ratio is located at the velocity edges of GMCs.
Since the velocity dispersion of the outer envelope
of a molecular cloud tends to be larger than that of the interior
(e.g. \cite{sakamoto2003}),
the results indicate that GMCs in this region are
heated from the outside of the clouds and
the outer envelope becomes hotter than the inside.
Detailed analysis will be shown in a forthcoming paper.
Although the higher ratio at the velocity edges is also seen
in the LV-diagram of $R_{21}$,
we suspect that this is artificially caused
by the substantial difference
of apparent velocities of the hyperfine lines;
19.3 km s$^{-1}$ for $(1,1)$ and 25.7 km s$^{-1}$ for $(2,2)$.

\begin{figure}[htbp]
 \begin{center}
  \FigureFile(160mm,210mm){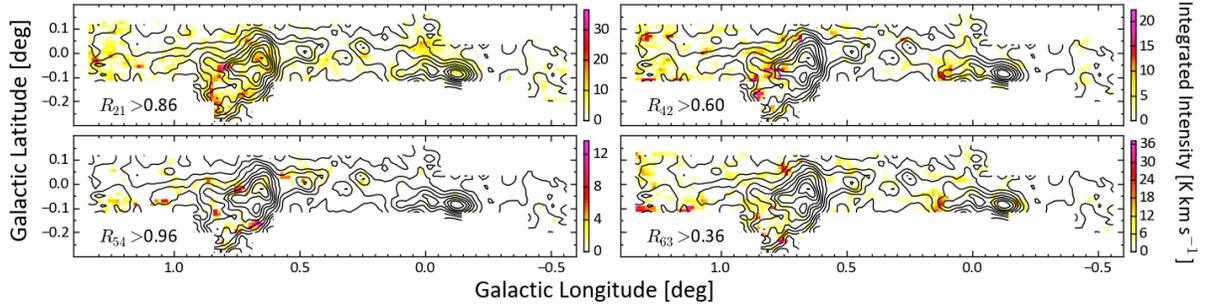}
 \end{center}
 \caption{
 Integrated intensity maps using only voxel
 where the intensity ratio is in the HR (color)
 overlaid on the distributions of the integrated intensity
 of the $(J,K)=(2,2)$ line (contours).           
 $T_{\rm mb}$ of the lower transitions are integrated.
 }\label{fig:high_ratio_map}
\end{figure}

Figure \ref{fig:high_ratio_map} shows integrated intensity maps
of the lower transitions $T_{\rm mb}(l,l)$ of the four intensity ratios
using only voxels in the HR
(i.e., voxel with $R_{21}>0.86$, $R_{42}>0.60$,
$R_{54}>0.96$, and $R_{63}>0.36$).
These maps look quite different from one another.
We consider that
this is mainly due to the difference of contribution of cold gas component
and $R_{54}$ is the most reliable indicator of warm gas tracer
because of the highest excitation energy of the lower transition.
The bright region in the $R_{54}$ map in figure \ref{fig:high_ratio_map}
indicates regions where extremely high-temperature
($T_{\rm rot} >300$ K) molecular gas exists.
For example, there is a bright region in the south of the Sgr-B2 region,
at $(L,B)=(\timeform{0D.69},\timeform{-0D.15})$.
This is fairly close to the region where
Tsuboi et al. (2015) argued a cloud-collision event.

\subsection{Comparison with Data of Other Molecules}

\begin{figure}[htbp]
 \begin{center}
  \FigureFile(150mm,70mm){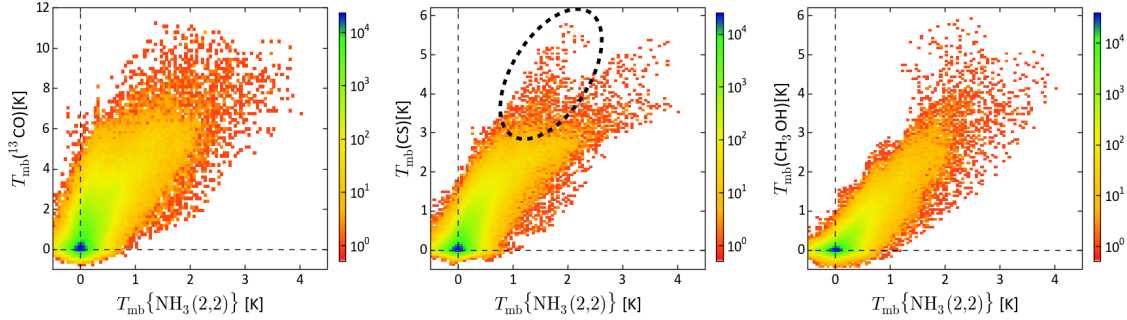}
 \end{center}
 \caption{
 $T_{\rm mb}$ scatter plot between NH$_3$ $(2,2)$ and
 other molecules in the observed area;
 {\it left}) $^{13}$CO ($1$--$0$) (Oka et al. 1998),
 {\it middle}) CS ($1$--$0$) (Jones et al. 2013),
 {\it right}) CH$_3$OH ($1_0$--$0_0$) (Jones et al. 2013).
 The color indicates number of voxels.
 }\label{fig:NH3vsCS}
\end{figure}

To grasp the physical condition and the environment of molecular gas,
we compared our NH$_3$ intensity with some other molecular line data
at same positions.
Figure \ref{fig:NH3vsCS} shows scatter plots between $T_{\rm mb}$
of our NH$_3$ $(2,2)$ and $^{13}$CO (1--0) (\cite{oka1998}),
CS (1--0), and CH$_3$OH ($1_0$--$0_0$) (\cite{jones2013}).
NH$_3$ $(2,2)$ is suitable for the comparison for the following reason;
(1) contribution of both of the two temperature components
(subsection \ref{sec:blp}),
(2) the relative intensities of the hyperfine lines
(to the main line) are not strong (table \ref{tab:nh3}).
These data were summed into each 2 km s$^{-1}$ velocity bin
and the velocity range was cut to be $V_{\rm LSR}=-200$ to $200$ km s$^{-1}$
to fit our data.
The spatial resolutions were also equalized to our data by smoothing.
The original intensities of CS and CH$_3$OH were in the $T_{\rm A}^*$ scale.
We converted them to the $T_{\rm mb}$ scale using
$\eta_{\rm mb}\approx0.43$ (\cite{jones2013}).

In contrast to $^{13}$CO,
CS shows a better correlation with NH$_3$,
which stemmed from the difference of the critical densities, $n_{\rm cr}$;
NH$_3$ and CS ($J=1$--$0$) trace dense molecular gas
($n_{\rm cr}>10^4$ cm$^{-3}$),
while $^{13}$CO ($J=1$--$0$) trace less dense molecular gas
($n_{\rm cr}=10^{2\mathchar`- 3}$ cm$^{-3}$).
We found the intensity ratio of NH$_3$ to CS to be low
in the majority of voxels of majority of the Sgr-A complex region
(except of 20--km s$^{-1}$ cloud, GCM$-$0.13$-$0.08).
The voxels are located around
$\{T_{\rm mb}({\rm NH}_3), T_{\rm mb}({\rm CS})\}=(1.8,5.0)$
(enclosed by an ellipse in figure \ref{fig:NH3vsCS}).
This indicates a strong UV environment in the region,
because CS is one of the molecules most tolerant
to such a region (\cite{drdla1989})
while NH$_3$ is a UV-fragile molecule (\cite{lee1984}).
As NH$_3$,
most of poly-atomic molecules such as CH$_3$OH, HNCO, and HC$_3$N
are UV-fragile and
weak around the Sgr-A complex region (\cite{jones2012}).
Actually, CH$_3$OH shows the tightest correlation among them,
which suggests similarity of creation/dissociation processes
of NH$_3$ and CH$_3$OH.
More detailed analysis will be shown in the forthcoming paper.

\section{Summary}

We carried out a survey of NH$_3$ $(J,K)=(1,1)$--$(6,6)$
in the major part of the CMZ with the Tsukuba 32-m telescope.
\begin{enumerate}
 \item Significant emission (S/N $>3$) of the NH$_3$ $(1,1)$--$(3,3)$ lines
       was simultaneously detected in 87\% of the observed area
       (2323 out of 2655 positions).
       Among the 2323 positions,
       NH$_3$ $(4,4)$--$(6,6)$ were also detected with S/N $>3$
       at 1426, 1150 and 1359 positions, respectively.
 \item The distribution and intensity
       of the NH$_3$ $(1,1)$ and $(2,2)$ data
       were consistent with the previous survey data
       obtained with the Mopra 22-m telescope.
 \item Some Boltzmann plots at representative points are shown.
	    There seems to be two temperature components.
 \item Typical intensity ratios
       among the NH$_3$ lines were $R_{21}=0.71$, $R_{42}=0.45$, $R_{54}=0.65$,
       and $R_{63}=0.17$.
       The distribution of $R_{63}$ tends to be higher
       at outer parts of giant molecular clouds (GMCs) than at inner parts.

\end{enumerate}

\bigskip

{\it Acknowledgements.}
We are very grateful to the VLBI group of the GSI
for licensing to use the 32-m telescope.
The observations have been made under the agreement
on the collaboration between the University of Tsukuba and the GSI.
We thank Yuki SUKETA, Yoji MORI, Miyabi TSUBOKAWA,
Kei-ichi ZEMPO, Shigeki TAKAOKA, Seika TAKAYANAGI, Takuya SUENAGA,
Takao ICHIHARA, Shota FUKUOKA, Yuki OKURA, Sumire AOKI, Hirokazu MASUDA,
Kazush KITAGAWA, Tsuguru RYU, Shoko KITAMOTO, Kazuki KOBAYASH,
and Hiroki OKUTOMI
for participating in the construction of the K-band observing system
of the Tsukuba 32-m telescope.
This work was supported in part by JSPS
(Japan Society for the Promotion of Science)
KAKENHI Grant Numbers 173400052, 17654042 and 20244011.



\begin{thebibliography}{}
\bibitem[Ao et al. (2013)]{ao2013}
Ao, Y., Henkel, C., Menten, K., Requena-Torres, M.A., Stanke, T., Aalto, S., M\"{u}hle, S., \& Mangum, J.
2013, \aap, 550, 125
\bibitem[Binney et al. (1991)]{binney1991}
Binney, J., Gerhard, O.E., Stark, A.A., Bally, J. \& Uchida, K.I. 1991, \mnras, 252, 210
\bibitem[Danby et al. (1988)]{danby1988}
Danby, G., Flower, D.R., Valiron, P., Schilke, P. \& Walmsley, C.M. 1988, \mnras, 235, 229
\bibitem[Drdla et al. (1989)]{drdla1989}
Drdla, K., Knapp, G.R., \& Dishoeck, E.F. 1989, \apj, 345, 815
\bibitem[de Pater et al. (2005)]{depater2005}
de Pater, I., DeBoer D., Marley M., Freedman, R. \& Young R. 2005, ICARUS, 173, 425
\bibitem[Flower et al. (1995)]{flower1995}
Flower, D.R., Pineau des For\^ets, G. \& Walmsley, C.M. 1995, \aap, 294, 815
\bibitem[Fukui et al. (2006)]{fukui2006}
Fukui, Y., et al. 2006, Science, 314, 106
\bibitem[Gordon \& Sorochenko (2002)]{gordon2002}
Gordon, M.A. \& Sorochenko, R.L., 2002, {\it Radio Recombination Lines}, Springer
\bibitem[Goldsmith \& Langer (1999)]{goldsmith1999}
Goldsmith, P.F. \& Langer, W.D. 1999, ApJ, 517, 209
\bibitem[Goto et al. (2008)]{goto2008}
Goto, M., Usuda, T., Nagata, T., Geballe, T.R., McCall, B.J., Indriolo, N., Suto, H., Henning, T., Morong, C.P. \& Oka, T. 2008, \apj, 688, 306
\bibitem[G\"{u}sten et al. (1981)]{guesten1981}
G\"{u}sten, R., Walmsley, C.M. \& Pauls, T. 1981, \aap, 103, 197
\bibitem[Hasegawa et al. (1994)]{hasegawa1994}
Hasegawa, T., Sato, F., Whiteoak, J.B. \& Miyawaki, R. 1994, \apj, 429, 77
\bibitem[Honma et al. (2012)]{honma2012}
Honma, M. et al. 2012, PASJ, 64, 136
\bibitem[H\"{u}ettemeister et al. (1993)]{huett1993}
H\"{u}ttemeister, S., Wilson, T.L., Bania, T.M. \& Mart\'in-Pintado, J. 1993, \aap, 280, 255
\bibitem[Jones et al. (2012)]{jones2012}
Jones, P.A., Burton, M.G., Cunningham, Requena-Torres, M.A., Menten, K.M.,
Schilke, P., Belloche, A., Leurini, S., Mart\'in-Pintado, J., Ott, J. \& Walsh, A.J. 2012, \mnras, 419, 2961
\bibitem[Jones et al. (2013)]{jones2013}
Jones, P.A., Burton, M.G., Cunningham, M.R., Tothill, N.F. \& Walsh, A.J. 2013, \mnras, 433, 211
\bibitem[Koyama et al. (2007)]{koyama2007}
Koyama, K., et al. 2007, \pasj, 59, 245
\bibitem[Kukolich (1967)]{kukolich1967}
Kukolich, S.G. 1967, Phys. Rev., 156, 83		
\bibitem[Kutner \& Ulich (1981)]{kutner1981}
Kutner, M.L. \& Ulich, B.L. 1981, \apj, 250, 341		
\bibitem[Lee (1984)]{lee1984}
Lee, L.C. 1984, \apj, 282, 172
\bibitem[Lis et al. (2010)]{lis2010}
Lis, D.C., Phillips, T.G., Goldsmith, P.F. et al., 2010, \apj, 521, L26
\bibitem[McGary \& Ho (2002)]{mcgary2002}
McGary, R.S. and Ho, P.T.P. 2002, \apj, 577, 757
\bibitem[Menten et al. (2009)]{menten2009}
Menten, K.M., Wilson, R.W., Leurini, S. \& Schilke, P. 2009, \apj, 692, 47
\bibitem[Mills \& Morris (2013)]{mills2013}
Mills, E.A.C. \& Morris, M.R. 2013, \apj, 772, 105
\bibitem[Morris et al. (1983)]{morris1983}
Morris, M., Polish, N., Zuckerman, B., \& Kaifu, N. 1983, \apj, 88, 1228
\bibitem[Morris \& Serabyn (1996)]{morris1996}
Morris, M., \& Serabyn, F. 1996, \araa, 34, 645
\bibitem[Nagayama et al. (2007)]{nagayama2007}
Nagayama, T., Omodaka, T., Handa, T., Iahak, H.B.H., Sawada, Y., Miyaji, T. \& Koyama, Y.
2007, \pasj, 59, 869
\bibitem[Nagayama et al. (2009)]{nagayama2009}
Nagayama, T., Omodaka,T., Handa, T., Toujima, T., Sofue, Y., Sawada, T., Kobayashi, H. \& Koyama, Y.
2009, \pasj, 61, 1023
\bibitem[Oka et al. (1998)]{oka1998}
Oka, T., Hasegawa, T., Sato, F., Tsuboi, M. \& Miyazaki, A. 1998, \apj, 118, 455
\bibitem[Pierce-Price et al. (2000)]{pierce2000}
Pierce-Price, D., et al. 2000, \apj, 545, L121
\bibitem[Purcell et al. (2012)]{purcell2012}
Purcell, C.R.., Longmore, S.N., Walsh, A.J., Whiting, M.T., Breen, S. L., Britton, T.,
Brooks, K. J., et al. 2012, \mnras, 426, 1972
\bibitem[Riquelme et al. (2013)]{riquelme2013}
Riquelme, D., Amo-Baladr\'on, M.A., Mart\'in-Pintado, J., Mauersberger, S., Mart\'in, S., \& Bronfman, L. 2013, \aap, 549, A36
\bibitem[Rodriguez-Fernandez et al. (2001)]{rod2001}
Rodr\'iguez-Fern\'andez, N.J., Mart\'in-Pintado, J. Fuente, A., de Vicente, P., Wilson, T.L. \& H\"{u}ttemeister, S.
2001, \aap, 365, 174
\bibitem[Sakamoto \& Sunada (2003)]{sakamoto2003}
Sakamoto, S. \& Sunada, K. 2003, \apj, 594, 340
\bibitem[Simmons \& Gordy (1948)]{simmons1948}
Simmons, J. W. \& Gordy, W. 1948, Phys. Rev., 73, 713
\bibitem[Tanaka et al. (2009)]{tanaka2009}
Tanaka, K., Oka, T., Nagai, M. \& Kamegai, K. 2009, \pasj, 61, 461
\bibitem[Tsuboi et al. (1999)]{tsuboi1999}
Tsuboi, M., Handa, T. \& Ukita, N. 1999, \apjs, 120, 1
\bibitem[Tsuboi et al. (2015)]{tsuboi2015}
Tsuboi, M., Miyazaki, A., and Uehara, K. 2015, \pasj, 67, 90
\bibitem[Ulich \& Haas (1976)]{ulich1976}
Ulich, B.J. \& Haas, R.W. 1976, \apjs, 30, 247
\bibitem[Walmsley \& Ungerechts (1983)]{walmsley1983}
Walmsley, C.M. \& Ungerechts, H. 1983 \aap, 122, 164
\bibitem[Walsh et al. (2011)]{walsh2011}
Walsh, A.J., Breen, S.L., Britton, T., Brooks, K.J., Burton, M.G., et al. 2011, \mnras, 416, 1764
\bibitem[Wilson et al. (1982)]{wilson1982}
Wilson, T.L., Ruf, K., Walmsley, C.M., Mart\'in, R.N., Batrla, W. \& Pauls, T.A. 1982 \aap, 115, 185
\bibitem[Wilson et al. (2006)]{wilson2006}
Wilson, T.L., Henkel, C. \& H\"{u}ettemeister, S. 2006 \aap, 460, 533
\bibitem[Yusef-Zadeh et al. (2007)]{yusef2007}
Yusef-Zadeh, F., Muno, M., Wardle, M. \& Lis, D.C. 2007, \apj, 656, 847

\end{thebibliography}
\end{document}